\documentclass[twocolumn,showpacs,preprintnumbers,amsmath,amssymb]{revtex4}

\usepackage{graphicx}% Include figure files
\usepackage{dcolumn}% Align table columns on decimal point
\usepackage{bm}% bold math

\begin{document}
\title{Violation of the Minimum H-H Separation ``Rule'' for Metal Hydrides}
\author{P.~Ravindran}
\email{ponniah.ravindran@kjemi.uio.no}
\homepage{http://folk.uio.no/ravi}
\author{P.~Vajeeston}
\author{R.~Vidya}  
\author{A.~Kjekshus} 
\author {H.~Fjellv{\aa}g}
\affiliation{Department of Chemistry, University of Oslo, Box 1033, 
Blindern, N-0315, Oslo, Norway}
\date {\today}
\begin{abstract}
Using gradient-corrected, all-electron, full-potential, density-functional 
calculations, including structural relaxations, it is found
that the metal hydrides $RT$InH$_{1.333}$ ($R$ = La, Ce, Pr, or Nd;
$T$ = Ni, Pd, or Pt)
possess unusually short H$-$H separations.
The most extreme value (1.454\,{\AA}) ever obtained for metal hydrides
occurs for LaPtInH$_{1.333}$. 
This finding 
violates the empirical rule for metal hydrides, which states that the minimum H$-$H
separation is 2\,{\AA}. 
Electronic structure, 
charge density, charge transfer, and electron localization 
function analyses on $RT$InH$_{1.333}$ show dominant metallic bonding with 
a non-negligible ionic component between $T$ and H, the H$-$H interaction being
weakly metallic. 
The paired, localized, and bosonic nature of the electron distribution at the H site
are polarized towards La and In which reduces the
repulsive interaction between negatively charged H atoms. This could explain the
unusually short H$-$H separation in these materials.
Also, $R$$-$$R$ interactions contribute to shielding of
the repulsive interactions between the H atoms. 
\end{abstract}
\pacs{71., 81.05.Je, 71.15.Nc, 71.20.-b}
\maketitle
The most attractive aspect of metal hydrides from a technological point
of view is their potential use as energy-storing materials.
Hydrogen as energy carrier and other visions of the ``hydrogen
society'' are especially attractive from an
environmental point of view, but since hydrogen is a  low-density gas
at STP, the storage of the large quantities required for most applications is a challenge.
High-pressure-compressed-gas storage is energy intensive if
high volume efficiency is desired, liquid or solid hydrogen storage even
more so, and all involve certain hazards. Storage of hydrogen in the form of solid metal hydrides, from which it
can readily be recovered by heating, is safe and volume
efficient.
\par
The amount of hydrogen per volume unit in metal hydrides is very high; in some cases higher
than in liquid or solid hydrogen, e.g., VH$_2$ stores more than twice the amount of hydrogen
than solid H$_2$ at 4.2\,K.
It is unfortunate, however, that most metal hydrides are heavy in relation
to the amount of hydrogen they contain. 
FeTiH$_2$ and LaNi$_5$H$_7$, e.g., only contain 1.9 and
1.6\,wt.$\%$ hydrogen, respectively. Therefore, efforts in hydride research have
over the past 25$-$30 years been concentrated on designing new, or modifying 
known, intermetallic
hydrides to increase the storing capacity and simultaneously adjusting
their properties to make them capable of delivering hydrogen
at useful pressures ($>$0.1 MPa) and acceptable temperatures ($<$ 425 K)~\cite{maelan01}.
These aspects are particularly important for most mobile applications where
hydrogen would be used directly in combustion engines or indirectly via fuel cells. 
It has proven
difficult to exceed 2\,wt.$\%$ of stored hydrogen and it remains a challenge
to increase this figure if metal hydrides are to become a viable  
source for the transportation sector. 

\par
The search for efficient hydrogen-storage metal hydrides~\cite{schlapbach94}
has to some extent been hampered by the mental barriers which empirical rules have
put on the thinking.  
For example, the interstitial hole size where hydrogen is
expected to occupy should be $>$~0.40~{\AA}. Switendick~\cite{switendick79}
observed from a compilation of experimental structure data that the minimum
H$-$H separation in ordered metal hydrides is $>$2{\AA} (``the 2-{\AA} rule''). 
This empirical pattern is later~\cite{rao85} supported
by band-structure calculations which ascribe the effect to repulsive 
interaction generated by the partially charged hydrogen atoms. A practical consequence
of this repulsive H$-$H interaction in metal hydrides is that it puts a limit 
to the amount of hydrogen which can be accommodated within a given structural framework. 
So, if H$-$H separations less than 2\,{\AA} would be possible
this could open for new efforts to identify potential intermetallics for higher hydrogen storing capacity.
However, there are indeed metal hydrides which do violate
``the 2-{\AA} rule'' and we have here identified the origin for such behavior.
\par
\begin{figure}
\includegraphics[scale=0.5]{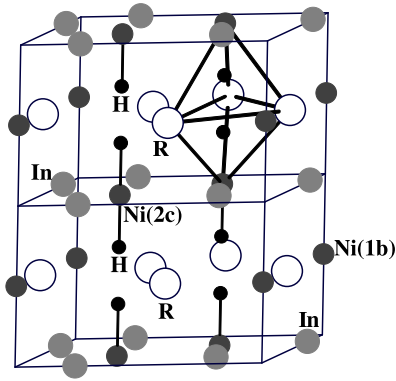}
\caption{\label{fig:str} The $R$NiInH$_{1.333}$-type crystal structure. The
$R_{3}$Ni$_2$ bipyramid is emphasized with thicker lines.} 
\end{figure}

$R$NiIn ($R$ = La, Ce, Pr, and Nd) crystallizes in the ZrNiAl-type structure (space 
group ${P\overline 62m}$) 
and can formally be considered as a layered arrangement with a
repeated stacking of two different planar nets of composition $R_{3}$Ni$_2$ and
NiIn$_3$ along [001] of the hexagonal unit cell.
If the hydrides of these materials obey the hole-size demand and ``the 
2-{\AA} rule'' one would
expect H to occupy the interstitial $2d$ site within $R_{3}$Ni$_2$ trigonal
bipyramid.
However, proton magnetic resonance (PMR) studies 
suggest~\cite{ghoshray92,sen96}
that H occupies both the 4$h$ and 6$i$ sites or either of them 
with H$-$H distances in the range 1.5$-$1.8 {\AA}.
%% ravi
%This lead to the local hydrogen density around eight times greater than in
%liquid hydrogen.
Recent powder X-ray and neutron diffraction studies~\cite{yartys} on $R$NiInD$_{1.333-x}$ 
(ideally $R_{3}$Ni$_3$In$_3$D$_4$)
show that deuterium occupies the 4$h$ site located on three-fold axis of 
$R_{3}$Ni tetrahedra that share a common face to form trigonal 
bipyramid (Fig.\,\ref{fig:str}).
This configuration gives rise to extraordinary short H$-$H separations of around 1.6 
{\AA}~\cite{yartys}.
As the diffraction techniques generally determine the average structure, 
neglect of partial H-site occupancies 
and local lattice distortions may lead one to conclude with
shorter H$-$H separations than actually present in the real 
structure~\cite{yvon88}. Hence, it is  
of interest to perform structural optimization theoretically. 
%%%%%%%%%%%%%%%%%%%%%%%%% FIG.1 %%%%%%%%%%%%%%%%%%%%%%%%%%%%%%%%%%
%%%%%%%%%%%%%%%% FIGURE %%%%%%%%%%%%%%%%%%%

\par
The present full-potential LMTO~\cite{wills} calculations 
are all electron, and no shape approximation to the charge density or
potential has been used. 
The basis functions, charge density and potential were expanded in spherical harmonic
series inside the muffin-tin spheres and in Fourier series in the interstitial regions.
The ratio of interstitial to unit cell volume is around 0.42.
The
calculations are based on the generalized-gradient-corrected-density-functional
theory as proposed by Perdew {\em et al.}~\cite{pw96} Spin-orbit terms
are included directly in the Hamiltonian matrix elements
for the part inside the muffin-tin spheres.
The basis set contained semicore $5p$ and valence $5d$,
$6s$, $6p$, and $4f$ states for La
(for Ce the $4f$ electrons are treated alternatively as valence and localized core
electrons, whereas the Nd-$4f$ and Pr-$4f$ electrons are treated as localized electrons using open core
approximation),
$3s$, $3p$, $4s$, $4p$, and $3d$ for Ni, $4s$, $4p$, $5s$, $5p$, and $4d$ for Pd, $5s$, $5p$, $6s$, $6p$, and $5d$ for Pt,
$5s$, $5p$, and $5d$ for In and $1s$, $2p$, and $3d$ states for H.
All orbitals were contained in the same energy panel.
To ensure well-converged wave functions a so-called
multi basis was included, implying the use of
different Hankel or Neuman functions each attaching to its own
radial functions. This allows reliable description of the
higher-lying unoccupied states and low-lying semicore states.
The self consistency was obtained with 105 {\bf k}
points in the irreducible part of the Brillouin zone.
Test calculations were made for 210 {\bf k} points to check for convergence, but 
the optimized
$c/a$ ratio for 105 and 210 {\bf k} points for LaNiInH$_{1.333}$ is
essentially the same.
To gauge the bond strength and nature of bonding we have used crystal orbital 
Hamiltonian population (COHP) and electron localization function (ELF)
analyzes, as is implemented in TBLMTO-47~\cite{cohp}.
\par
All calculations relate to ideal and fully saturated hydrides with composition
$R_{3}T_{3}$In$_3$H$_4$ ($RT$InH$_{1.333}$; $R$ = La, Ce, Pr, or Nd, $T$ = Ni, Pd, or Pt).
For (trivalent) $R$ the
$4f$ electrons were treated as core electrons (except for La; 4$f$). As Ce-$4f$ electrons
are known to take different valence states in intermetallic compounds,
different possibilities~\cite{delin98} for the valence states of                           
Ce were considered in both the hydrides and intermetallic phases during
the structural optimization.
For this
optimization all atom positions were relaxed by force minimization and 
equilibrium $c/a$ and volume were obtained by total energy minimization. 
Optimized structural parameters for the Ce compounds are in good agreement
with experimental values (only) when Ce atoms are assumed to be in the trivalent state.
The calculated equilibrium lattice parameters and the changes between the intermetallic 
and corresponding hydride phases
are given for selected compounds along with experimental parameters
in Table \ref{table:latparam}. 
%%%%%%%%%%%%%%%%%%%%%%%%% Table 1 end %%%%%%%%%%%%%%%%%%%%%%%%%%%%%%%%%%
\begin{table*}
\caption{Calculated lattice parameters ($a$ and $c$ in \AA) and $c/a$ for La$T$InH$_{1.333}$ 
and relative variation in unit cell dimensions (in $\%$) consequent on hydrogenation
from $RT$In to $RT$InH$_{1.333}$.}
\begin{ruledtabular}
\begin{tabular}{|l|c|c|c| c |c| c|c |c|c |c |c |c|}
 & \multicolumn{2}{c|} {$a$} & \multicolumn{2}{c|}{$c$ } &
 \multicolumn{2}{c|} {$c/a$ } & \multicolumn{2}{c|}{$\Delta a/a$ } &
 \multicolumn{2}{c|} {$\Delta c/c$ } & \multicolumn{2}{c|}{$\Delta V/V$ }\\ \cline{2-3}\cline{3-5}
\cline{5-7}\cline{7-9}\cline{9-11}\cline{11-13}
 Compound         & The.   & Exp.   & The.   & Exp.   & The.   & Exp.   & The.   & Exp.  & The.  & Exp. & The. & Exp.  \\ \hline
LaNiInH$_{1.333}$ & 7.2603 & 7.3810 & 4.5522 & 4.6489 & 0.6270 & 0.6399 & $-$3.969 & $-$2.76 & 14.02 & 14.8 & 5.14 & 8.54 \\
LaPdInH$_{1.333}$ & 7.3501 & ---    & 4.8112 & ---    & 0.6546 & ---    & $-$5.42  & ---   & 16.64 &  --- & 4.33 & ---   \\
LaPtInH$_{1.333}$ & 7.7274 & ---    & 4.6903 & ---    & 0.6070 & ---    & $-$0.04  & ---   & 13.98 &  --- & 14.00 & --- \\ 
CeNiInH$_{1.333}$ & 7.4536 & 7.2921 & 4.4871 & 4.6238 & 0.6020 & 0.6341 & $-$1.68  & $-$3.21 & 12.72 & 16.3 & 8.97 & 8.98  \\
PrNiInH$_{1.333}$ & 7.3783 & 7.260  & 4.4726 & 4.560  & 0.6062 & 0.6281 & $-$2.85  & $-$3.73 & 13.93 & 15.4 & 7.52 & 7.01 \\
NdNiInH$_{1.333}$ & 7.2408 & 7.2255 & 4.5560 & 4.5752 & 0.6292 & 0.6332 & $-$3.72  & $-$3.92 & 16.75 & 16.5 & 7.60 & 7.53  \\
\end{tabular}
\end{ruledtabular}
\label{table:latparam}
\end{table*}
%%%%%%%%%%%%%%%%%%%%%%%%% Table 1 end %%%%%%%%%%%%%%%%%%%%%%%%%%%%%%%%%%
%Owing to the lack of space we have not included 
%results for the pure $RT$In phases and some of the $RT$InH$_{1.333}$ phases in the listing.
\par
In general, the calculated lattice parameters are
in good agreement with the experimental values, and the small differences
found may partly be attributed to hydrogen
non-stoichiometry (around 10$\%$) in the experimental studies.
The hydrogen-induced lattice expansion is strongly anisotropic
(Table \ref{table:latparam}); 
a huge expansion along [001] ($\Delta c/c$ = 14$-$20\%) and 
a smaller intralayer
contraction ($-\Delta a/a$ = 0$-$5.8\%).
The calculated cohesive energy and heat of formation for the hydrides are
larger than for the corresponding intermetallic phases indicating that it might be possible to
synthesize all these hydrides. The electronic structure studies 
show that all considered phases are in the metallic
state consistent with experimental findings. 
The calculated $R-$H, $T-$H, and H$-$H distances are
given in Table\,\ref{table2} along with experimentally available values.
%%%%%%%%%%%%%%%%%%%%%%%%% Table 2 end %%%%%%%%%%%%%%%%%%%%%%%%%%%%%%%%%%
\begin{table}
\caption{Calculated  interatomic distances (in {\AA}) for $RT$InH$_{1.333}$.}
\begin{ruledtabular}
\begin{tabular}{|l|c|c|c|c|c|c|}
& \multicolumn{2}{c|} {$R-$H} & \multicolumn{2}{c|}{$T-$H } &\multicolumn{2}{c|} {H$-$H } \\ \cline{2-3}\cline
{3-5}\cline{5-7}
Compound & The.~~~~ & Exp. & The. & Exp. & The. & Exp. \\ \hline
LaNiInH$_{1.333}$ & 2.379 & 2.406 & 1.489 & 1.506 & 1.573   & 1.635 \\
LaPdInH$_{1.333}$ & 2.373 & --- & 1.644  & --- & 1.523    & --- \\
LaPtInH$_{1.333}$ & 2.475 & --- & 1.618 & --- & 1.454   & --- \\ 
CeNiInH$_{1.333}$ & 2.427 & 2.371 & 1.457 & 1.508 & 1.572   & 1.606 \\
PrNiInH$_{1.333}$ & 2.387 & --- & 1.492 & --- & 1.487  & --- \\
NdNiInH$_{1.333}$ & 2.350 & 2.350 & 1.493 & 1.506 & 1.492   & 1.562   \\
\end{tabular}
\end{ruledtabular}
\label{table2}
\end{table}
%%%%%%%%%%%%%%%%%%%%%%%%% Table 1 end %%%%%%%%%%%%%%%%%%%%%%%%%%%%%%%%%%
An interesting observation 
is that all $RT$InH$_{1.333}$ materials have unusually short H$-$H distances.
Two explanations have been proposed.
Pairing of the hydrogen atoms (either by molecular H$_2$-like bonding 
or by bonding mediated by the intermediate $T$ atom)
has been advanced to explain 
the anomalous PMR spectrum of CeNiInH$_{1.0}$~\cite{ghoshra93}.
The second explanation focuses on the significantly shorter
La$-$La distance in LaNiInH$_{1.333}$ 
than in closely related phases~\cite{yartys}, whereby the La atoms (generally
$R$) may act as a
shielding that compensates the repulsive H$-$H interaction.
\par
\begin{figure}
\includegraphics[scale=0.45]{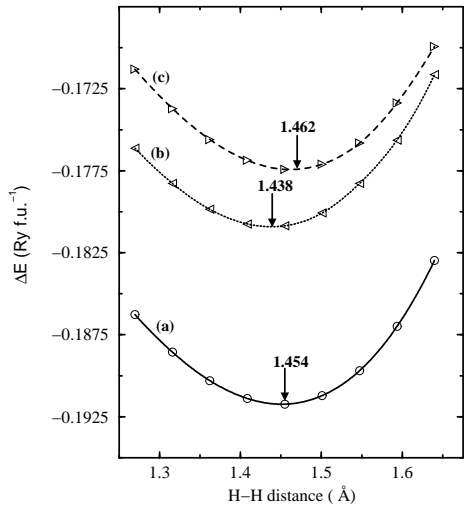}
\caption{\label{fig:hh} Total energy versus H$-$H distance in LaPtInH$_{1.333}$. (a) All atoms except
H are fixed at their equilibrium positions.(b) La atoms are moved 0.08 {\AA} out of their
equilibrium position toward H.
(c) La atoms are moved 0.08 {\AA} out of their equilibrium position away from H.}
\end{figure}

In order to evaluate these possibilities we have calculated the
total energy for (hypothetical) LaPtInH$_{1.333}$ as a function of H$-$H separation according
to three different scenarios: 
(1) Keeping La, Pt, and In fixed in their equilibrium positions,
(2) moving La 0.08 {\AA} out of the equilibrium position toward H, and
(3) moving La 0.08 {\AA} out of the equilibrium position away from H. 
The obtained results are illustrated in Fig.\,\ref{fig:hh}.  
When La, Pt, and In are in their optimized equilibrium position, 
the equilibrium H$-$H separation is 1.454\,{\AA}. This scenario
corresponds to a lower total energy than the two alternatives. For scenario 3 
we obtain a shorter H$-$H separation (1.438 {\AA})
than for the ground state configuration, and for scenario 2 a correspondingly larger
separation (1.462~{\AA}). 
%%%%%%%%%%%%%%%%%%%%%%%%% FIG.2 %%%%%%%%%%%%%%%%%%%%%%%%%%%%%%%%%%
%%%%%%%%%%%%%%%% FIGURE %%%%%%%%%%%%%%%%%%%
\par
As the total energy curves increase steadily on  reduction of
the H$-$H separations, the possibility of stabilization of hydrogen in the form
of molecular H$_2$-like units seems completely ruled out. 
The total energy increases drastically also for increased H$-$H separation beyond 
the equilibrium value. 
This is due to  increasing
repulsive $T-$H interaction and decreasing attractive H$-$H interaction. 
{\it The considerable changes in the equilibrium H$-$H distance on $R$ 
displacement indicate that $R$ in the $R_{3}T_{2}$ trigonal bipyramidal configuration (Fig.\,\ref{fig:str})
acts as a shielding that to some extent compensates repulsive H$-$H interactions}.
When the $R-$H separation is reduced the H
atoms are allowed to approach each other more closely. 
\par
Owing to charge transfer from metal to
hydrogen the repulsive H$-$H interaction in metal hydrides are
generally larger than that within the H$_2$ molecule and this may be the physical
basis for ``the 2-{\AA} rule''. 
Another example
of violation of ``the 2-{\AA} rule'' is found for Th$_2$AlH$_4$~\cite{vaji1}, but here
the minimum H$-$H separation (1.945\,{\AA}) within the Th$_5$ trigonal bipyramids is much larger than
the bond distance in the H$_2$ molecule and the H$-$H separations in $RT$InH$_{1.333}$.
In order to quantify the bonding interaction between the constituents in the
$RT$InH$_{1.333}$ series the integrated crystal orbital Hamilton population
(ICOHP) were calculated. For example, the ICOHP values up to
$E_{F}$ for LaNiInH$_{1.333}$ are
$-$3.44, $-$0.14, $-$0.72, $-$0.85, $-$0.86, $-$1.21, and $-$0.61\,eV for
Ni(2c)$-$H, H$-$H, La$-$H, Ni(2c)$-$In, La$-$Ni(1b), Ni(1b)$-$In, and Ni(2c)$-$La,
respectively. This
indicates that the strongest bonds are those between Ni(2c)
and H. Another important observation 
is that the bonding interaction between the hydrogens is small, which further
confirms that   
the short H$-$H separation in these materials are not rooted in hydrogen pairing or 
formation of H$_2$-like molecular units.
%%%%%%%%%%%%%%%%%%%%%%%%% FIG.3 %%%%%%%%%%%%%%%%%%%%%%%%%%%%%%%%%%
\begin{figure*}
\includegraphics[width=\textwidth]{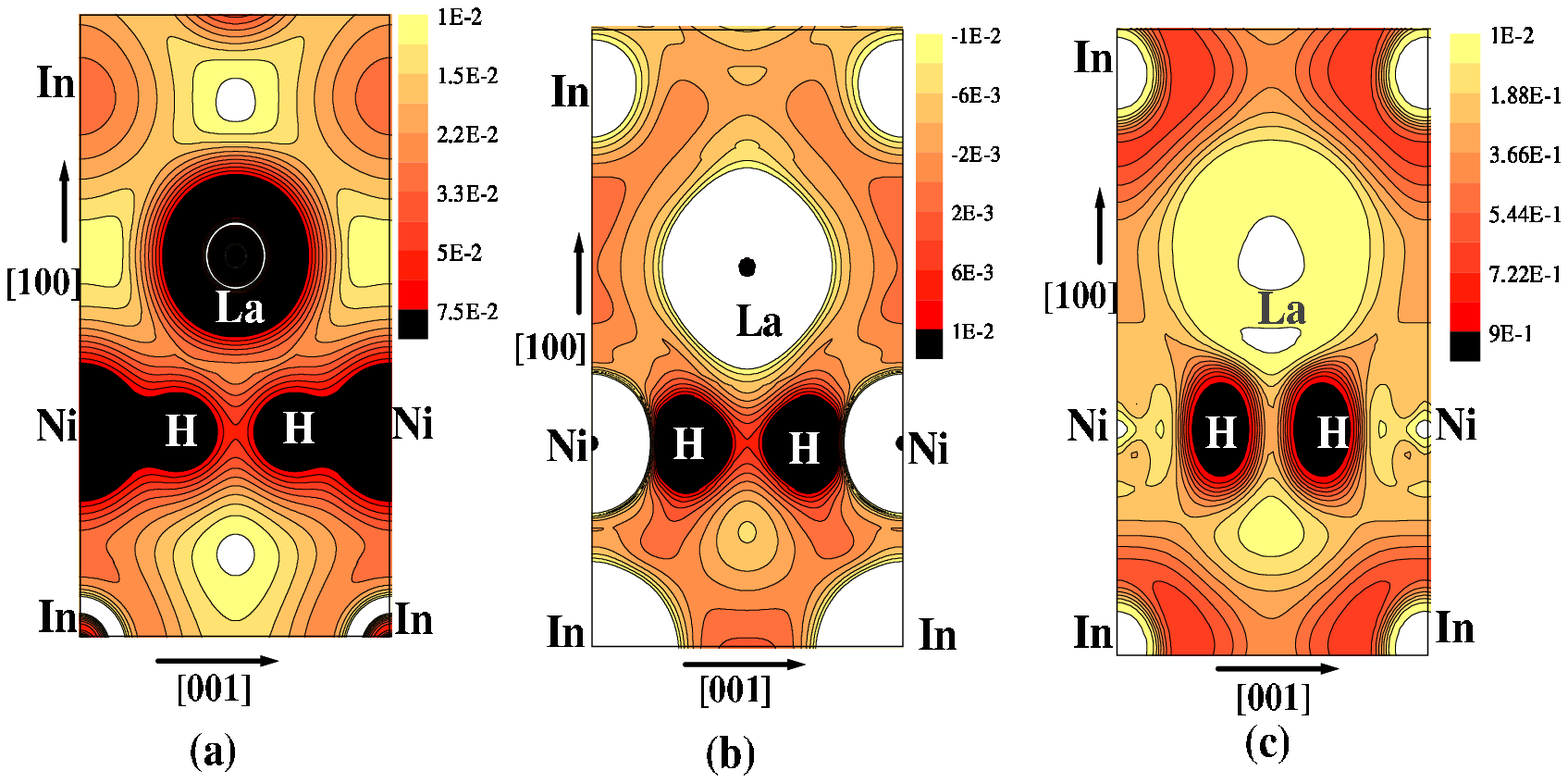}
\caption{\label{fig:charge} (a) Total charge density, (b) Charge transfer and (c) Electron localization
function plot for LaNiInH$_{1.333}$ in the 
(100) plane. The origin is shifted to (1/3, 0, 0) and the charge densities are in e/a.u.$^{3}$.}
\end{figure*}
%%%%%%%%%%%%%%%%%%%%%%%%% FIG.3 %%%%%%%%%%%%%%%%%%%%%%%%%%%%%%%%%%
\par
In order to substantiate this observation further we have calculated the 
valence-charge-density distribution in (100) of LaNiInH$_{1.333}$ 
(Fig.\,\ref{fig:charge}a). From this
figure, it is apparent that Ni and H form an NiH$_2$ molecule-like
structural subunit. Moreover, Fig.\,\ref{fig:charge}a demonstrates that there is 
no substantial charge density distributed between
the H atoms.
In order to depict the role of charge transfer 
we have displayed the charge transfer (the difference in the electron
density of the compound and that of constituent atoms superimposed on the
lattice grid) for LaNiInH$_{1.333}$ within (100)  
in Fig.~\ref{fig:charge}b. 
From Fig.~\ref{fig:charge}b it is clear that electrons are transfered
from La, In, and Ni to the H site. So, there is considerable ionic bonding component between H
%% ravi
and the metallic host lattice. 
The transferred electrons from the metallic host lattice to the H$_2$-like subunit
of the structure enter the antibonding $\sigma^{*}$ levels and gives rise to
repulsive interaction. This repulsive interaction between the negatively
charged H atoms could explain why the H-H separation in these materials are
larger than that in the H$_2$ molecule. If there was strong covalent bonding between Ni
and H one should expect a significant (positive) value of charge transfer distribution
between these atoms
(contributed by both atoms). The absence of such a feature rules out this 
possibility.
The ELF is an informative tool
to distinguish different bonding interactions in solids ~\cite{elf} and 
ELF for LaNiInH$_{1.333}$ in (100) is given in Fig.~\ref{fig:charge}c.
The large value of ELF at the H site indicates strongly paired
electrons with local bosonic character. Another manifestation of covalent 
bonding between Ni and H should have been paired electron distribution between these atoms.
The negligibly small ELF between Ni and H indicates that the probability of
finding parallel spin electrons close together is rather high (correspondingly
small for antiparallel spin pairs)
in this region confirm metallic bonding consistent with the result obtained from
charge transfer analysis and the detailed analysis show that delocalized
metallic Ni(2c)-$d$ electrons are distributed in this region.
Even though the charge distribution between Ni and H looks like a typical covalent
bonding the charge transfer and ELF analyzes clearly show that the electron
distribution between Ni and H are having parallel spin alignment and purely
from Ni site.  
Hence, chemical bonding between Ni and H is dominated by metallic components
with considerable ionic weft. The partial density of state analysis
also show that the H-$s$ states are well separated from
the Ni-$d$ states in the whole valence band indicates the presence of ionic 
bonding between Ni and H.
Due to the repulsive interaction between the negatively charged H electrons, the
ELF contours are not spherically shaped but polarized towards La and In. {\it The localized
nature of the electrons at the H site and their polarization towards La and In 
reduce significantly the H$-$H repulsive interaction and this can 
explain the unusually short H$-$H separation in this compound.}
The ELF between the H atom takes a significant value of 0.35.
Considering the small charge density, this indicates a weak 
metallic type of interaction between the hydrogen atoms.

\par
$RT$InH$_{1.333}$ constitutes a series with much shorter H$-$H separations than
other known metal hydrides.
We have shown that the short distances between the H atoms in such metal hydrides are governed primarily
by the polarization of negative charges on H towards the electropositive
La and In.
We believe that this
conclusion is of more general validity, and may be utilized to search for
other metal hydrides of potential interest as hydrogen storage materials.

\par
The authors are grateful to the Research Council of Norway for financial support
and for computer time at the Norwegian supercomputer facilities.

\end{document}